\documentclass[12pt]{article}

\usepackage{graphicx}

\newcommand{\beq}{\begin{equation}}
\newcommand{\eeq}{\end{equation}}
\newcommand{\bea}{\begin{eqnarray}}
\newcommand{\eea}{\end{eqnarray}}

\newcommand{\be}{\begin{equation}}
\newcommand{\ee}{\end{equation}}
\newcommand{\ba}{\begin{array}}
\newcommand{\ea}{\end{array}}

\newcommand{\La}{\Lambda}

   \newcommand{\CD}{\mathcal{D}}
   \newcommand{\CF}{\mathcal{F}}

   \newcommand{\CN}{\mathcal{N}}

\newcommand{\CW}{\mathcal{W}}

\newcommand{\br}[1]{\left( #1 \right)}

\newcommand{\dint}[2][]{\mathop{\mathalpha{\int#1}#2}}

\font\cmss=cmss10 at 11pt \font\cmsss=cmss8 at 8pt

\def\inbar{\vrule height1.5ex width.4pt depth0pt}
\def\mininbar{\vrule height.75ex width.3pt depth0pt}
\def\cc{\relax\,\hbox{$\mininbar\kern-.2em{\hbox{\rm\tiny C}}$}}
\def\IZ{\relax\ifmmode\mathchoice
{\hbox{\cmss Z\kern-.4em Z}}{\hbox{\cmss Z\kern-.4em Z}}
{\lower.4pt\hbox{\cmsss Z\kern-.4em Z}}
{\lower1.2pt\hbox{\cmsss Z\kern-.4em Z}}\else{\cmss Z\kern-.4em Z}\fi}
\def\IC{\relax\,\hbox{$\inbar\kern-.3em{\rm C}$}}
\def\IR{\relax{\rm I\kern-.18em R}}

\newcommand{\SU}{\mathrm{SU}}

\newcommand{\PP}{\mathrm{I}\kern -2pt \mathrm{P}}
\newcommand{\R}{\mathrm{I}\kern -2.5pt \mathrm{R}}
\newcommand{\Z}{\mathsf{Z}\kern -5pt \mathsf{Z}}
\newcommand{\1}{1\kern -3pt \mathrm{l}}

\newcommand{\tr}{{\rm tr}}
\newcommand{\Tr}{{\rm Tr}}
\newcommand{\pa}{\partial}

\newcommand{\cN}{\mathcal{N}}

\newcommand{\e}{{\rm e}}

\def\half{ {\textstyle\frac{1}{2}} }

\def\gs{ g_{s} }

\def\sumN{ \sum_{i=1}^{N} }

\def\vev#1{ \langle {#1} \rangle }

\begin{document}

\begin{flushright} 
{\tt hep-th/0405242}\\ 
BRX-TH-542\\ 
\end{flushright}
\vspace{1mm} 
\begin{center}
{\bf\Large\sf 
Exact Superpotentials, Theories with Flavor 
and Confining Vacua}

\vskip 5mm
Marta~G\'omez-Reino\footnote{Research 
supported by the DOE under grant DE--FG02--92ER40706.\\
{\tt \phantom{aaa} E-mail:\,marta@brandeis.edu}}\\

\end{center}

\vskip 1mm

\begin{center}
{\em Martin Fisher School of Physics\\
Brandeis University, Waltham, MA 02454}
\end{center}

\vskip 2mm

\begin{abstract} 

In this paper we study some interesting properties of the 
effective superpotential of $\CN=1$ 
supersymmetric gauge theories with fundamental matter, with 
the help of the Dijkgraaf--Vafa proposal connecting supersymmetric gauge 
theories with matrix models. 

We find that the effective superpotential 
for theories with $N_f$ fundamental flavors can be calculated in terms 
of quantities computed in the pure ($N_f=0$) gauge theory. 
Using this property we compute 
in a remarkably simple way the exact effective superpotential of $\CN=1$ 
supersymmetric theories with fundamental matter and gauge group $SU(N_c)$, 
at the point in the moduli space where a maximal number 
of monopoles become massless (confining vacua). We extend the analysis to a generic 
point of the moduli space, and show how to 
compute the effective superpotential in this general case.

\end{abstract}


\section{Introduction}
\setcounter{equation}{0}

Over the past few years much progress has been made in computing 
effective superpotentials of $\CN=1$ supersymmetric gauge theories. Motivated 
by geometric considerations of dualities in string theory \cite{civ,cv}, an 
expression for the quantum effective superpotential was proposed by Dijkgraaf 
and Vafa. They conjectured that the effective superpotential can be calculated 
by doing perturbative computation in an auxiliary matrix model \cite{dv,dv2,dv3}, 
being later proved with perturbative field theory arguments in 
\cite{c} and by the analysis of the generalized Konishi anomaly in 
\cite{witten,seiberg,witten2}. This proposal provides direct connections between 
the computations in the matrix model descriptions with those in 
supersymmetric gauge theories. 
The proposal was later extended to the addition of matter in the fundamental 
representation of the gauge group in \cite{arg,m1,m2,m3,howard2,hof}.

In this paper we will be considering the $\CN=1$ supersymmetric gauge theory 
with matter in the fundamental representation of the gauge group 
which can be obtained by deforming the $\CN=2$ gauge theory via the addition of a tree 
level superpotential. Actually there has been significant work on using the 
Dijkgraaf--Vafa proposal to get results of $\CN=2$ theories 
\cite{howard2,gukov,fer,fer2,im2,howard,yo}

In the case of $\CN=1$ theories with fundamental matter the 
effective superpotential will have contributions coming from planar diagrams with 
one boundary, apart from the contribution coming from the planar 
diagrams with no boundaries \cite{arg}. As we will see in the first part of the paper, 
those contributions to the superpotential 
can be computed, within the matrix model setup, in terms of traces of 
certain matrix model operators. We find that those traces should be computed in 
the pure gauge theory, even to calculate the effective superpotential for theories 
with flavor. As a direct 
application of this we will compute the exact superpotential for a $\CN=1$ theory 
with gauge group $SU(N_c)$ and $N_f<N_c$ fundamental matter hypermultiplets 
at the point of the moduli space where a 
maximal number of monopoles become massless. From the point of 
view of the underlying Seiberg--Witten curve 
\cite{sw} this correspond to the point where the curve factorizes completely. 
The moduli that factorizes the Seiberg--Witten curve in the pure gauge theory 
(no matter) case are well 
known \cite{ds}, and they will be the only ingredient that we need 
to compute the exact effective superpotential in this case, even though we are 
considering theories with fundamental matter. 
We will find that the result for the exact superpotential in this case is 
remarkably simple. We will also consider the generalization of these results 
to an arbitrary point of the moduli space with $n$ distinct glueball superfields, by 
using the techniques developed in \cite{witten} to compute traces of operators 
within matrix models.

The plan of this paper is as follows: in Section 1 we will review briefly the 
Dijkgraaf--Vafa proposal for $\CN=1$ theories with $SU(N_c)$ gauge group, 
and $N_f$ matter hypermultiplets in the fundamental representation of the 
gauge group, and we will set up the 
ingredients that we will need for the calculation of the effective 
superpotential. In Section 3 we will compute, using the results in 
Section 2, the exact effective superpotential for $\CN=1$ theories with 
fundamental matter in the point of the moduli space where a maximal number 
of monopoles become massless. In section 4 we will extend the analysis of 
Section 3 to a generic point of the moduli space and show how to 
compute the effective superpotential in this general case. Finally, in 
Section 5 we conclude with a summary of the work.

\section{The Effective Superpotential of ${\cal N}=1$ Supersymmetric 
Gauge Theories}

In this paper we will be considering an $\SU(N_c)$ gauge theory 
with $N_f<N_c$ pairs of quark fields 
$q_i$ ($\tilde q_i$), $i=1,\cdots,N_f$, in the fundamental  
(anti-fundamental) representation of 
$\SU(N_c)$. The Lagrangian density of this theory is given by 
\begin{equation}
{\cal L}= \dint{d^4\theta}\Tr(\bar q_i\e^V q_i+\tilde q_i\e^V\bar{\tilde q}_i)
   + \dint{d^2\theta}\Tr(W(\phi,q,\tilde q)+\tau \CW^\alpha \CW_\alpha),
\end{equation}
where $\CW_\alpha$ is the gauge superfield and the superpotential $W$ is given by 
\begin{equation}\label{pot}
  W(\phi,q,\tilde q) = W(\phi) + \phi q\tilde q - q m\tilde q.
\end{equation}
The first term of (\ref{pot}) is a polynomial tree level superpotential for the 
adjoint Higgs field $\phi$ and $m$ is the mass matrix for the flavors. 
Classically at the critical points of this tree level superpotential the gauge 
group breaks to $\prod_k SU(N_k)$ where $N_k$ is the number of eigenvalues of 
the Higgs field $\phi$ at the $k$-th critical point. Quantum mechanically 
there will be a gluino condensate $S_k$ for each of the factors and an 
effective superpotential for these condensates. In this section 
we will review the Dijkgraaf--Vafa approach to 
the computation of the effective superpotential for these condensates 
as a planar limit of a given matrix models \cite{dv,dv2,dv3}. 

\subsection{Theories With Flavors from Matrix Models}

According to the proposal given by Dijkgraaf and Vafa in \cite{dv,dv2,dv3} 
the effective superpotential of ${\cal N}=1$ gauge 
theories obtained as a deformation of ${\cal N}=2$   
theories by an arbitrary tree 
level superpotential can be computed using matrix models. 
In the generalization of this proposal to include fields in the 
fundamental representation of the gauge group (this case was first 
considered in \cite{arg}) this superpotential is given by
\beq\label{weff}
W_{eff}(S) = N_c\frac{\partial\CF_{\chi=2}}{\partial S}+ \CF_{\chi=1}\,,
\eeq
where the $\CF_{\chi=1,2}$ are defined through the matrix integral
\begin{equation}\label{eq}
  Z = \e^{-\sum_{\chi}\frac{1}{g_s^\chi}\CF_\chi} = \frac{1}{Vol(G)}
\dint{\CD\Phi\CD q\CD\tilde q}\e^{-\frac{1}{g_s}\Tr W(\Phi,q,\tilde q)}\,,
\end{equation}
and the superpotential $W(\Phi,q,\tilde q)$ is given by Eq.(\ref{pot}), 
but replacing the gauge theory fields by matrices $(\phi,q,{\tilde q})
\rightarrow (\Phi,Q,{\tilde Q})$. Following the approach 
in \cite{hof} for the generalization of the Dijkgraaf-Vafa 
proposal to theories with fundamental matter, $\Phi$ will be here  
a $M\times M$ matrix, $Q$ a $M\times M_f$ matrix and ${\tilde Q}$ a 
$M_f\times M$ matrix, where the parameters $M$, $M_f$ are unrelated to the 
gauge theory parameters $N_c$, $N_f$.  

Then, introducing the parameters $S\equiv g_sM$ and $S_f\equiv g_sM_f$
the dependence of the free energy on the quantities $g_s$, $M$, and 
$M_f$ can easily be extracted from the topology of the diagrams, and 
can be written as an expansion in the genus $g$ and the number of 
quark loops $h$ as
\begin{equation}\label{free}
  \CF = \sum_{g,h}g_s^{2g-2}S_f^h\CF_{g,h}(S). 
\end{equation}
The planar contributions can be found from a large $M$ expansion of the 
matrix model. All planar diagrams are summed by taking both the rank 
of the gauge group and the number of flavors to infinity 
($ M,\,M_f\to\infty$) and ($g_s\to 0$), while keeping $S= g_sM$ and 
$S_f= g_sM_f$ finite. From the expansion in (\ref{free}) we see that 
this limit picks out 
the genus zero contribution, and also that to reproduce the effective 
superpotential (\ref{weff}) one should consider at most one quark loop 
and set $S_f$ to zero at the end of the calculation (see \cite{hof} 
for details)
\begin{equation}\label{eff}
  W_{eff}(S) = \left. N_c\frac{\partial\CF}{\partial S}\right|_{S_f=0}+
\left. N_f\frac{\partial\CF}{\partial S_f}\right|_{S_f=0}\equiv
N_c\frac{\partial\CF_{\chi=2}}{\partial S} + \CF_{\chi=1}
\end{equation}

Furthermore, as the matter fields in (\ref{eq}) appear only quadratically, 
we can integrate them out. This generates a $\log (\Phi+m)$ 
potential for $\Phi$ and we are then left with an integral 
just over $\Phi$ \cite{m1,howard2,hof}
\begin{equation}\label{zeta}
  Z = \e^{-\CF}=\dint{\CD\Phi}\e^{-\frac{1}{g_s}\Tr {\tilde W}(\Phi)},
\end{equation}
where 
\begin{equation}\label{v}
  {\tilde W}(\Phi) = W(\Phi) + S_f\sum_{k=1}^{N_f}\log(\Phi-m_k)\,,
\end{equation}
and $W(\Phi)$ is an arbitrary tree level superpotential
\beq
W(\Phi)=\sum_{p\geq1}g_p\frac{\tr \Phi^p}{p}\,.
\eeq

In the following subsection we will show how to deduce some very interesting 
properties of $\CF_{\chi=2}$ and $\CF_{\chi=1}$ with the help of 
(\ref{free}), (\ref{zeta}) and (\ref{v}). Those properties will be  
actually useful to compute the effective superpotential exactly in 
several cases.

\subsection{Properties of the ${\cal F}_{\chi=2}$ and 
${\cal F}_{\chi=1}$ Contributions to the Superpotential}

The $\CF_{\chi=2}$ and $\CF_{\chi=1}$ contributions to the 
effective superpotential can 
be evaluated perturbatively about an extremal point of the tree level 
superpotential \cite{gukov,howard,howard2}. From the perturbative expansion it 
is easy to see that the form of $\CF_{\chi=2}$ in terms of the glueball 
superfield remains unchanged when including 
fundamental matter in the theory. All the explicit dependence in the matter 
of the hypermultiplets appears in the $\CF_{\chi=1}$ contribution. 

\vspace{.3cm}

\noindent$\bullet$ \underline{$\chi=2$ Contribution}

\vspace{.3cm}

Taking derivatives of the free energy with respect to 
the parameter $g_s$ has been proven to be a useful tool to get 
non--trivial information about the effective superpotential 
in the pure gauge group (no matter) case \cite{pestun}. We will check 
here that the relation obtained in \cite{pestun} (see also \cite{matone}) 
remains unchanged with the 
addition of matter and we will also see how to use that relation to 
compute the $\CF_{\chi=2}$ contribution to the superpotential.

From Eq.(\ref{free}) we can calculate the derivative of the free energy 
with respect to $g_s$ within our setup
\beq
\label{uno}
\pa_{g_s} \CF =  \sum_{g,h} g_s^{2g-3} (S_f)^h \left( (2g-2) \CF_{g,h} + 
S\frac{\pa \CF_{g,h}}{\pa S} \right)+
\sum_{g,h}hg_s^{2g-3}(S_f)^h\CF_{g,h}(S)\,.
\eeq

As we already mentioned in the previous section the genus zero 
contribution is obtained by taking the $g_s=0$ limit and 
by considering at most one quark loop ($S_f=0$). Therefore we get that 
in this limit
\beq
\label{dos}
\pa_{g_s} \CF =  g_s^{-3}\left(S\frac{\pa \CF_{\chi=2}}{\pa S} 
-2 \CF_{\chi=2} \right)\,.
\eeq
Note that the relation (\ref{dos}) does not involve any explicit 
contribution from the matter of the hypermultiplets, as it does not 
depend on $\CF_{\chi=1}$. 

On the other hand, since in the planar limit the partition 
function $Z$ of the matrix 
model is given by (\ref{zeta}), the free energy is
\beq\label{fe}
\CF =-\log Z= -\log \br{ \int d \Phi e^{-g_s^{-1} \tr {\tilde W}(\Phi)}}\,.
\eeq
From the definition of $\CF$ as a free energy 
we know that its derivative over a parameter
is a vacuum expectation value of the correspondingly coupled operator. 
Therefore
\beq\label{tres}
\pa_{g_s} \CF = -g_s^{-2} \vev { \tr {\tilde W}(\Phi)}+
g_s^{-2} S_f \sum_{k=1}^{N_f}\vev {\tr \log (\Phi+m_k)}=
-g_s^{-2} \vev{ \tr W(\Phi)}\,,
\eeq
where we have used the definition of $\tr {\tilde W}(\Phi)$ 
given by (\ref{v}). 
Remember that, as we already mentioned, we are interested on the limit that 
takes into account just the genus zero contribution (that is, $g_s=0$ and 
$S_f=0$). This means that, once we take that limit, the vacuum 
expectation value in (\ref{tres}) has to be computed as in the pure 
$SU(N_c)$ theory. Then, comparing (\ref{dos}) and (\ref{tres}) we get that
\beq\label{trw}
g_s\, \vev{ \tr W(\Phi)}_0=  2 \CF_{\chi=2}-
S \frac{\pa \CF_{\chi=2}}{\pa S}\,,
\eeq
where by $\vev{...}_0$ we mean vacuum expectation value 
in the pure $SU(N_c)$ theory. The relation (\ref{trw}) is exactly 
the same as the one found in \cite{pestun} for a gauge theory without 
matter content, and it is therefore a check that the form of 
$\CF_{\chi=2}$ in terms of the glueball superfield $S$ 
remains unchanged with the addition of matter.
Also the fact that within matrix models there is a well developed 
technique to compute vacuum expectation values \cite{witten} implies 
that Eq.(\ref{trw}) can be used to compute 
$\CF_{\chi=2}$ to all orders in perturbation theory. We will show 
this explicitly in the following sections.

\vspace{.3cm}

\noindent$\bullet$ \underline{$\chi=1$ Contribution}

\vspace{.3cm}

As we mentioned in the beginning of this section, all the matter dependence 
that will appear on the effective superpotential will be encoded on the 
$\CF_{\chi=1}$ contribution. As in the previous case, we would like 
to find an expression for $\CF_{\chi=1}$ involving vacuum expectation values, 
 but taking derivatives of the free energy with respect to the parameter $g_s$ 
does not give us any information about $\CF_{\chi=1}$. Nevertheless in our case, 
contrary to the pure case, we still have another parameter in the theory: 
 $S_f$. If we take a derivative of the free energy $\CF$ in (\ref{free}) with 
respect to $S_f$ we get 
\beq\label{sf}
\pa_{S_f} \CF=\sum_{g,h}hg_s^{2g-2}(S_f)^{h-1}\CF_{g,h}(S)\,.
\eeq
Therefore the genus zero contribution ($g_s=0$ and $S_f=0$) will 
pick out the term $\pa_{S_f} \CF = \frac{1}{g_s^2}\CF_{\chi=1}$. 
When taking a derivative of (\ref{fe}) with respect to $S_f$ we get
\beq\label{sf2}
\pa_{S_f} \CF =g_s^{-1}\sum_{k=1}^{N_f}\vev {\tr \log (\Phi+m_k)}\,.
\eeq
Therefore in the genus zero limit we have that (by comparing Eq.(\ref{sf}) 
and (\ref{sf2}))
\beq\label{f1}
\CF_{\chi=1} =g_s\sum_{k=1}^{N_f}\vev 
{\tr \log (\Phi+m_k)}_0\,,
\eeq
where the {\it vev} is again meant to be computed in terms of the pure $SU(N_c)$ theory. We
 will show in the following sections how can (\ref{f1}) be used to calculate 
$\CF_{\chi=1}$.

\section{Exact Superpotentials for Theories with Flavors in Confining Vacua}
\setcounter{equation}{0}

As a first application to the relations (\ref{trw}) and (\ref{f1}) obtained 
in the previous section we will compute in this section the exact effective 
superpotential of a certain ${\cal N}=1$ supersymmetric theory. We will 
consider a $\CN=1$ theory obtained by 
perturbing a ${\cal N}=2$ supersymmetric by a tree level superpotential 
$SU(N_c)$ gauge theory with $N_f$ matter hypermultiplets, 
in the point of the moduli space where 
$N_c-1$ monopoles become massless. The exact effective superpotential 
was computed in \cite{fer} for a 
pure gauge theory using the ``integrating in'' procedure. The addition of 
fundamental matter was considered in \cite{janik} using random matrix 
models. The method described in this section shows that in order  
to obtain the exact superpotential with 
fundamental matter in the maximally degenerating point of the moduli 
space we just need to compute, within the matrix model formulation, 
vacuum expectation values of several operators. This is a considerable advantage 
compared to previous works. As we will see, those 
vacuum expectation values are very easy to compute in the present case, 
due to the fact that we just need information about the pure 
gauge theory, so the actual computation of the exact effective superpotential 
turns out to be very simple.

\subsection{Exact Superpotentials for $SU(N_c)$ Theories in Confining Vacua}

The quantum moduli space of ${\cal N} = 2$ supersymmetric
$SU(N_c)$ gauge theories with $N_f$ massive flavors is $(N_c-1)$-dimensional, 
and it is parametrized by 
the moduli $u_k$, $k=2,\cdots,N_c$. At each point of the moduli space, 
the low energy theory
is described by an ${\cal N} = 2$ effective theory where the gauge group is 
broken to $U(1)^{N_c-1}$. All the information about hte $\CN=2$ theory 
is encoded in a particular meromorphic 
one--form $d\lambda_{SW}$ defined over an auxiliary curve, the 
Seiberg--Witten curve \cite{sw,SUN}
\beq
\label{curva}
y^2 = P_{N_c} (x, u_k)^2 - 4 \Lambda^{2N_c - N_f} \prod_{f=1}^{N_f}(x + m_f),
\eeq
where $P_{N_c} (x, u_k)$ is the characteristic polynomial of 
$SU(N_c)$ that is given by
\beq 
P_{N_c} (x, u_k) = x^{N_x}-\sum_{k=2}^{N_c}u_{k} x^{N_c-k}. 
\eeq
In this section we will be interested in the case where $N_c-1$ mutually local 
monopoles condense. 
This corresponds to a complete factorization of the Seiberg-Witten
curve
\beq\label{matter}
P_{N_c} (x, u_k)^2 - 4 \Lambda^{2N_c - N_f} \prod_{i=1}^{N_f}(x + m_i) =
(x-x_1)(x-x_2) H_{N_c-1}^2(x)
\eeq

For the case of pure $SU(N_c)$ the solution to this problem was found 
in \cite{ds} with the help of 
Chebyshev polynomials. The moduli that factorize the curve in 
the pure gauge theory case are given by
\begin{equation}\label{mod}
u_{2p}^0=\frac{N_c}{2 p}C_{2p}^p\La^{2p}\,,\,\hspace{2cm}
\,u_{2p+1}^0=0\,,
\end{equation}
where the $C_{2p}^{p}$ are the binomial 
coefficients \footnote{The case of $U(N_c)$ can be obtained from the case 
of $SU(N_c)$ by shifting $x\rightarrow x - u_{1}/N$. This will induce a 
shift in the moduli $u_p$ \cite{fer} that should be taken into account 
for the generalization of our results to the $U(N_c)$ theory.}. The generalization 
of (\ref{mod}) to the case with matter has been addressed in \cite{janik,warner,janik2}, 
and the expressions for the moduli that factorize the curve (\ref{matter}) become 
very complicated compared to the pure gauge theory case. However, for the computation 
of the exact effective superpotential that we develop in this section, we just 
need the simple form of (\ref{mod}).

One can deform this ${\cal N} = 2$ theory to a ${\cal N} = 1$ gauge 
theory by adding a tree level superpotential
\beq
\label{tree}
W_{tree} = \sum_{p\geq 1} \frac{g_p}{p}\, \tr \phi^p.
\eeq
The presence of this superpotential will lift the quantum moduli space,
characteristic of the
${\cal N} = 2$ Coulomb phase,  except for the 
dimension $1$ submanifolds, where $N_c-1$ mutually local magnetic
monopoles become massless \cite{sw}.

In order to compute the exact superpotential in this case it is useful 
to write Eq.(\ref{trw}) in a slightly different way. Let us take the 
derivative with respect to $S$, so that we get
\beq\label{ttrw}
\frac{\pa}{\pa S}\,g_s\, \vev{ \tr W(\Phi)}_0=  
\frac{\pa \CF_{\chi=2}}{\pa S}-S \frac{\pa^2 \CF_{\chi=2}}{\pa S^2}\,.
\eeq
Now we have all the information that we need to compute 
$\CF_{\chi=2}$ to all orders in perturbation theory. 
It was found in \cite{civ} that the full contribution coming from 
${\cal F}_{\chi=2}$ to the effective superpotential is given by
\beq\label{wo}
W_{eff}^0=N_c\,\frac{\pa \CF_{\chi=2}}{\pa S}=N_c \, S\, (-\log 
\frac{S}{\tilde \Lambda^3}+1)+N_c\frac{\pa \CF_{\chi=2}^{pert}}{\pa S}\,,
\eeq
where the first piece is the Veneciano-Yankielowitz superpotential for 
pure $SU(N_c)$ super Yang--Mills \cite{vy}. Also $\CF_{\chi=2}^{pert}$ 
is given by a perturbative expansion in $S$
\beq\label{fpert}
\CF_{\chi=2}^{pert}=\sum_{n\geq 1}f_n^{\chi=2}(g_p)\, S^{n+2}\,.
\eeq

Inserting (\ref{wo}) and (\ref{fpert}) in (\ref{ttrw}) we get
\beq\label{cc}
N_c\, \frac{\pa}{\pa S}\, g_s\, \vev{ \tr W(\Phi)}_0=  
W_{eff}^0-S \frac{\pa W_{eff}^0}{\pa S}=
N_c\, S-N_c \sum_{n\geq1}n(n+2)f_n^{\chi=2}(g_p)\, S^{n+1}\,.
\eeq

We should take now into account the following fact: when one considers 
a ${\cal N}=2$ theory, the moduli $u_n$ are given by 
$u_n = {1 \over n} \tr (\phi^n)$, where $\phi$ is the scalar component 
of the adjoint 
$\cN=1$ chiral superfield of the $\cN=2$ vector multiplet.
In the Seiberg-Witten approach, the {\it vevs} of these operators
may be written in terms of integrals over the cycles of the Seiberg--Witten 
curve. On the other hand, on the matrix model side the expression for 
the {\it vevs} of the moduli $u_n$ was calculated in \cite{howard,howard2}, 
and for the case we 
are considering in this section they are given by
\beq\label{utr}
u_p^0 ={N_c}{\pa \over \pa S} {\gs \over p} \vev{\tr( \Phi^p)}_0\,.
\eeq
As in the $W_{tree}=0$ case we should recover the $\cN=2$ theory, so that the 
$u_p$ in (\ref{utr}) should be the same ones as the $u_p$ in 
(\ref{mod}). Therefore
\beq\label{fin}
{\pa \over \pa S} g_s\, \vev{\tr( \Phi^{2p})}_0=
 C_{2p}^p\La^{2p}
\eeq
Using (\ref{fin}) it is easy to  see that at the critical point of the 
superpotential $\pa_SW_{eff}^0=0$ we get from (\ref{cc})
\beq\label{w0}
\sum_{p\geq1}g_{2p} u_{2p}^0=W_{eff}^0\,,
\eeq
as implied by the Intriligator, Leigh and Seiberg linearity principle 
\cite{ils,fer}. Also from here we can extract the relation between 
the glueball superfield $S$ and the scale of the theory $\Lambda$ at 
the critical point.
\beq\label{ese} 
S_0=\frac{1}{N_c}\frac{\pa W_{eff}^0}{\pa \log \Lambda^2}=
\sum_{p\geq1}\frac{1}{2}g_{2p}C_{2p}^p\La^{2p}\,.
\eeq
This is the same relation as found in \cite{fer}

Now at the critical point it is straight forward to obtain from (\ref{cc}) and (\ref{fin}) 
that 
\beq\label{exp}
\sum_{p\geq1}\frac{1}{2 p}g_{2p}C_{2p}^p\La^{2p}=S_0-\sum_{n\geq1}n(n+2)
f_n^{\chi=2}(g_p)\, S_0^{n+1}\,.
\eeq

Using (\ref{ese}) we are able to extract from (\ref{exp}) the 
coefficients $f_n^{\chi=2}(g_p)$ in a recursive way. Actually, we get 
the following expression for the $\CF_{\chi=2}$ to the effective superpotential
\bea\label{coef}
W_{eff}^0&=&N_c\,\frac{\pa \CF_{\chi=2}}{\pa S}=N_c S(-\log \frac{S}{g_2 \Lambda^2}+1)+
N_c\sum_{n\geq 1}(n+2) f_n^{\chi=2}(g_p) S^{n+1}\,,\\
f_1^{\chi=2}\,&=&\frac12\frac{g_4}{g_2^2}\nonumber\,,\\
f_{n\geq2}^{\chi=2}\,&=&
\frac{C_{2(n+1)}^{n+1}}{2(n+2)(n+1)}\frac{g_{2(n+1)}}{g_2^{n+1}}-
\sum_{l=1}^{n-1}\frac{l(l+2)}{n(n+2)}
\, f_l^{\chi=2}\hspace{-.5cm}\sum_{\stackrel{\scriptstyle p_1,\cdots,p_{l+1}=1}
{\scriptstyle p_1+\cdots+p_{l+1}=n+1}}^{n+1} \hspace{-.5cm}
\frac{C_{2p_1}^{p_1}g_{2p_1}\cdots 
C_{2p_{l+1}}^{p_{l+1}}g_{2 p_{l+1}}}{2^{l+1}\, g_{2}^{n+1}}\nonumber\,.
\eea

It is very easy to check that the exact result for the effective 
superpotential obtained here, and given by Eq.(\ref{coef}) coincides with the 
one obtained in \cite{fer}. One 
just has to substitute the expression for the exact effective superpotential 
in \cite{fer} in the Eq.(\ref{wo}) and check that it is fulfilled once 
one takes into account Eq.(\ref{fin}) and Eq.(\ref{ese}). 
The main difference in both expressions for the exact superpotential is that 
in \cite{fer} the dependence of the superpotential in the glueball 
superfield $S$ is given in an implicit way, whereas here we write 
that dependence in an explicit way. The results presented here also 
agree with the ones appearing in \cite{ook,georgi} for the special 
cases of quadratic and quartic tree level superpotentials. The expression 
(\ref{coef}) is valid for an arbitrary tree level superpotential. The 
result in (\ref{coef}) is just one part of the effective superpotential. 
To get the full answer we need now to compute the matter contribution 
encoded in $\CF_{\chi=1}$.

\subsection{The Matter Contribution}

Following the same guidelines as the ones presented in the previous 
subsection for the computation of $W_{eff}^0$, we can compute the exact 
form of the $\CF_{\chi=1}$ contribution to the effective superpotential. 

Expanding (\ref{f1}) around the critical point at $\Phi=0$ we have that
\beq\label{a}
\CF_{\chi=1}=\sum_{f=1}^{N_f}\left(\vev{\tr \log m_f}_0+
\sum_{k=1}^\infty\frac{(-1)^{k+1}}{k\, m_f^k}\vev{\tr\, \Phi^k}_0\right)\,.
\eeq

If we now take the derivative of (\ref{a}) with respect to the glueball 
superfield $S$ we find that
\beq\label{f1ds}
\frac{\pa \CF_{\chi=1}}{\pa S}=\sum_{f=1}^{N_f}\left(\log m_f-
\sum_{k=1}^\infty\frac{C_{2k}^k}{2k}\frac{\Lambda^{2k}}{{m_f^{2k}}}\right)\,,
\eeq
where we have used the relation (\ref{fin}). Therefore $\CF_{\chi=1}$ 
is just given by 
\beq\label{int}
\CF_{\chi=1}=\int\sum_{f=1}^{N_f}\left(\log m_f-
\sum_{k=1}^\infty\frac{C_{2k}^k}{2k}\frac{\La^{2k}}{m_f^{2k}}\right)\left(
\frac{\pa S_0}{\pa \Lambda}\right)d\Lambda
\eeq

Then, using (\ref{ese}) we can perform the integral in ({\ref{int}) so 
that we get
\beq\label{f1def}
\CF_{\chi=1}=\sum_{f=1}^{N_f}\sum_{l\geq1}\frac12g_{2l}C_{2l}^l\Lambda^{2l}
\, \log m_f-\sum_{f=1}^{N_f}
\sum_{k,l\geq1}\frac{lC_{2l}^lC_{2k}^kg_{2l}}
{4k(k+l)m_f^{2k}}\La^{2(k+l)}
\eeq

If now we want the explicit dependence in the glueball superfield $S$ 
of $\CF_{\chi=1}$ in (\ref{f1def}) we can do it by 
using (\ref{ese}), so that again we can compute the coefficients 
$f_n^{\chi=1}(g_p,m_f)$ recursively
\bea\label{f1def2}
\CF_{\chi=1}&=&S\sum_{f=1}^{N_f}\log m_f+
\sum_{n\geq1}f_n^{\chi=1}(g_p,m_f)S^{n+1}\,,\hspace{1.5cm}
f_1^{\chi=1}=-\frac12\sum_{f=1}^{N_f}\frac{1}{g_2m_f^2}\,,\\
f_{n\geq2}^{\chi=1}&=&-\frac{1}{g_2^{n+1}}\sum_{f=1}^{N_f}\left(
\sum_{\stackrel{\scriptstyle k,l=1}{\scriptstyle k+l=n+1}}^{n}\frac{lC_{2l}^l
C_{2k}^{k}g_{2l}}{4k(n+1)m_f^{2k}}+
\sum_{q=1}^{n-1}\frac{f_q^{\chi=1}}{2^{q+1}}\hspace{-.5cm}\sum_{\stackrel
{\scriptstyle p_1,\cdots,p_{q+1}=1}{\scriptstyle p_1+\cdots+p_{q+1}=n+1}}^{n+1} 
\hspace{-.5cm}C_{2p_1}^{p_1}g_{2p_1}\cdots C_{p_{q+1}}^{p_{q+1}} g_{2 p_{q+1}}\right)
\nonumber\,.
\eea
Note that the form of the effective superpotential is additive with 
respect to inclusion of flavors.

Then the final expression for the exact effective superpotential, 
showing explicitly the dependence on the glueball superfield $S$ for 
an arbitrary tree level superpotential, is given by
\beq
W_{eff}=W_{eff}^0+\CF_{\chi=1}\,,
\eeq
where $W_{eff}^0$ is given in (\ref{coef}) and $\CF_{\chi=1}$ 
in (\ref{f1def2}). The 
fact that in the results presented in \cite{janik} the dependence on the 
glueball superfield enters in a highly non--linear way in the equation 
for the effective superpotential, makes it 
very complicated to compare both results. Nevertheless, we can check our 
result by comparing it with the one obtained in \cite{arg} (see also 
\cite{gripaios}) for the particular case of 
a quadratic tree level superpotential, $W_{tree}=\frac12\tr \Phi^2$. For 
this particular case we have that $g_2=1$, $g_n=0$, $n\neq2$. Then we just get 
from (\ref{coef}) and (\ref{f1def2})
\beq\label{eq6}
W_{eff}=N_c\,S\,(-\log\frac{S}{\Lambda^2}+1)+S\sum_{f=1}^{N_f}\log m_f-\sum_{f=1}^{N_f}
\sum_{k,l\geq1}\frac{(2k-1)!}
{k!(k+1)!m_f^{2k}}S^{k+l}\,.
\eeq
As it can be seen from (\ref{eq6}) we get the same perturbative expansion 
as the one obtained in \cite{arg} by a different procedure (they sum 
over all the planar diagrams using the results in \cite{planar}).

It deserves to be emphasized that the results given in (\ref{coef}) 
and (\ref{f1def2}) gives us the exact effective superpotential for an 
arbitrary level superpotential in a remarkably simple way. 
Also note that the dependence of the effective superpotential 
on the glueball superfield is written explicitly.

\section{Effective Superpotentials in General Vacua}
\setcounter{equation}{0}

In this section we will explain how to compute the effective superpotential 
using (\ref{trw}) and (\ref{f1}) at a general point of the moduli space of 
the gauge theory. Let us consider that we have a $\cN=2$ supersymmetric 
gauge theory broken to $\cN=1$ by the addition of a tree-level
superpotential $W(\phi)$ to the gauge theory
\beq
W_{tree}=\sum_{p=1}^{n+1}g_p\frac{\tr \phi^p}{p}\,.
\eeq 
Let us also consider that this tree level superpotential has $n$ non coincident 
critical points. This will mean that we will have $n$ glueball superfield $S_i$, 
$i=1,\cdots n$ (one at each critical point). 

Now we want to use in this case the analysis developed in Section 2 to compute the effective 
superpotential. For this purpose we have to take into account the fact that 
the formula (\ref{trw}) will now assume the following form
\beq\label{trw2}
g_s\, \vev{ \tr W(\Phi)}_0=  2 \CF_{\chi=2}-
\sum_{i=1}^{n}S_i \frac{\pa \CF_{\chi=2}}{\pa S_i}\,,
\eeq
according to the fact that that we have $n$ distinct glueball superfields. 
On the other hand, 
the formula (\ref{f1}) remains unchanged as does not involve 
derivatives with respect to $S$. 

In the case with $n$ glueball superfields, the form of $\CF_{\chi=2}$ 
around a critical point located at $e_i$ can be written as
{\small\beq
\label{free2}
\CF_{\chi=2} =  \sum_{i=1}^n S_i W(e_i) 
- \half \sum_{i=1}^n S_i^2 \log \left( S_i\over W''(e_i) \La^2 \right) 
- \sumN \sum_{j\neq i}  S_i S_j \log\left(e_i-e_j \over \La \right) + 
\sum_{p \geq 3} \CF_{p}^{\chi=2}
\eeq}
where the Veneziano--Yankielowitz part of $\CF_{\chi=2}$ was calculated 
in \cite{howard} from the matrix model integral. The coefficients $\CF_{m}^{\chi=2}$ 
are polynomials of order $p$ in $S_i$. Now introducing (\ref{free2}) into (\ref{trw2}) we get 
\beq\label{trw3}
g_s \, \vev{ \tr W(\Phi)}=  \sum_{i=1}^nS_iW(e_i)+\frac12 
\sum_{i=1}^nS_i^2-\sum_{m\geq3}(m-2)\CF_{m}^{\chi=2}\,.
\eeq
Also for the matter contribution we have
\beq\label{efe1}
\CF_{\chi=1} =\sum_{f=1}^{N_f}\sum_{i=1}^nS_i\,\log(e_i+m_k)+\sum_{q\geq2}\CF_q^{\chi=1}=g_s\sum_{f=1}^{N_f}\vev 
{\tr \log (\Phi+m_f)}\,,
\eeq
where $\CF_q^{\chi=1}$ are polynomials in $S_k$ of order $q$. The coefficients $\CF_{m}^{\chi=2}$ 
and $\CF_q^{\chi=1}$ can be computed perturbatively within the matrix model. However  
we now explain how to compute those coefficients just by using the spectral curve associated with 
the matrix model.

Therefore, in order to compute the effective superpotential we need to compute 
the vacuum expectation values in 
(\ref{trw3}) and (\ref{efe1}). Within the matrix model framework these 
expectation values can be calculated easily following \cite{witten} by 
introducing the resolvent
\beq
\omega(x) \equiv g_s \tr \vev {\frac 1 {x - \Phi}}\,.
\eeq

This resolvent is specified in terms of the loop equation 
\beq
\label{loop}
\omega(x)^2 = \omega(x) W'(x) + \frac 1 4 f_{n-1}(x)\,,
\eeq
where $f_{n-1}(z)$ is an arbitrary polynomial of the order $n-1$, and 
$W'(x)=\prod_{i=1}^n(x-e_i)$. 
From the equation (\ref{loop}) we read that the resolvent is given by
\beq
\omega(x) = \frac12 \left( W'(x) - \sqrt{W'(x)^2+f_{n-1}(x)}\right)\,.
\eeq
The values of the glueball superfields at the critical point $e_i$ will be 
specified by the  $n$ coefficients of the polynomial $f_{n-1}(x)$ through the 
integrals \cite{civ}
\beq
\label{esei}
S_i = \frac 1 {2\pi i} \oint_{A_i} \omega(x) dx\,,
\eeq
where by $A_i$ we denote a cycle enclosing the branch point centered in 
point $e_i$ of the curve $y=\sqrt{W'(x)^2+f_{n-1}(x)}$, where $y$ is the 
spectral curve associated with the matrix model.

Via the resolvent $\omega(x)$ one can easily calculate the expectation 
values of the single trace operators $\tr \Phi^k$ like \cite{witten}
\beq\label{vev}
g_s\vev{\tr \Phi^k} =  \frac 1 {2\pi i} \oint_{A} x^k\omega(x) dx
\eeq
where $A=\sum_{i=1}^n A_i$.

In order to being able to compute vacuum expectation values with the help of 
(\ref{vev}) we will use the parametrization of the polynomial $f_{n+1}$ that appears 
in (\ref{loop}) given in \cite{im}, that we have found to be very useful, and that is 
given by
\beq\label{fn}
f_{n-1}(x) = \sum_{i=1}^n \tilde S_i \prod_{j\neq i}^n(x-e_j) =
W'(x)\sum_{i=1}^n \frac{\tilde S_i}{x-e_i}\,.
\eeq
Note that this polynomial is of degree $n-1$ only, as it should be. 
Now with the help of (\ref{fn}) we can compute $S_i$ in terms of ${\tilde S}_k$ and 
$e_j$ by computing the period integral (\ref{esei}). This period integral can 
be computed by reducing the evaluation of the integral to a set of residue calculations. 
Therefore if we expand the resolvent (\ref{loop}) around the point ${\tilde S}_k$, 
$k=1,\cdots,n$, the integral (\ref{esei}) over the cycle $A_i$ can be performed 
just by calculating the residues at the point $x=e_i$
{\small\bea\label{s2}
&&S_i= \frac 1 {2\pi i} \oint_{A_i} \omega(x) dx=\\
&&={\tilde S}_i+\sum_{\stackrel{\scriptstyle p=0}
{m\geq2}}^m\,2^{m-2}\,\frac{(2m-3)!!}{p!(m-p)!}
\frac{{\tilde S}_i^p}{(m+p-2)!}\frac{\pa^{m+p-2}}
{\pa x^{m+p-2}}\frac{1}
{R_i(x)^{m-1}}\left(\sum_{j\neq i}\left.
\frac{{\tilde S}_j}{x-e_j}\right)^{m-p}
\right|_{x=e_i}\nonumber\,,
\eea}
where by $R_i(x)$ we denote the polynomial $R_i(x)=\prod_{j\neq i}(x-e_j)$. 
Note that at first order in ${\tilde S}$ we have that $S_i={\tilde S}_i$. 
This will be important later on. Also with the help of (\ref{fn}) and 
(\ref{vev}) we can compute $g_s\, \vev{\tr W(\Phi)}$ using the same procedure. 
We get that
{\small\bea\label{eq1}
&&g_s \vev{\tr W(\Phi)}=\frac 1 {2\pi i} \oint_{A} W(x)\omega(x) dx =
 \sum_{i=1}^nW(e_i){\tilde S}_i+\\
&&\left.+\sum_{\stackrel{\scriptstyle p=0}
{m\geq2}}^m\,2^{m-2}\,\frac{(2m-3)!!}{p!(m-p)!}
\frac{{\tilde S}_i^p}{(m+p-2)!}\frac{\pa^{m+p-2}}
{\pa x^{m+p-2}}\frac{W(x)}
{R_i(x)^{m-1}}\left(\sum_{j\neq i}\frac{{\tilde S}_j}{x-e_j}\right)^{m-p}
\right|_{x=e_i}\nonumber\,.
\eea}
Now, due to the fact that at first order on ${\tilde S}$ we have that $S_i={\tilde S}_i$, 
it is easy to see to use (\ref{s2}) to rewrite (\ref{eq1}) in the form ({\ref{trw3}), 
and extract from there the form of the coefficients $\CF_p^{\chi=2}$. For example 
for $\CF_3^{\chi=2}$ we get
{\small\beq
\CF_3^{\chi=2}=\sum_{i=1}^n\frac{S_i^3}{R_i^2}\left(\frac14R_i''-
\frac{2(R_i')^2}{3R_i}\right)+\frac{S_i^2}{R_i}\sum_{j\neq i}S_j\left(\frac{1}{e_{ij}^2}+
\frac{2R_i'}{R_ie_{ij}}\right)-\frac{2S_i}{R_i}\sum_{j\neq i, k\neq i,j}\frac{S_jS_k}
{e_{ij}e_{ik}}\,,
\eeq}
where $e_{ij}=e_i-e_j$ and $R_i=\prod_{j\neq i}(e_i-e_j)$, and we have used that 
$\pa^nW(e_i)=(n-1)\pa^{n-2}R_i(e_i)$. The result for 
$\CF_3^{\chi=2}$ agrees with the one 
computed in \cite{im} using Whitham hierarchies, and also with the one computed in \cite{howard} 
from a perturbative matrix model point of view. Also note that in the particular case where 
$S_{j\neq i}=0$ (that is, at the maximally degenerating point) we recover the 
results computed in the previous section for the coefficient of order three in $S$.

The coefficients $\CF_p^{\chi=2}$ give us one part of the effective superpotential. In 
order to get the matter contribution we need to compute (\ref{efe1}). Following the 
same procedure as before we get that
{\small\bea
&&g_s \sum_{f=1}^{N_f}\vev{\tr \log(\Phi+m_f)}=
\sum_{f=1}^{N_f}\frac 1 {2\pi i} \oint_{A} \log(x+m_f)\omega(x) dx =
\sum_{f=1}^{N_f}\sum_{i=1}^nS_i\log(e_i+m_f)+\nonumber\\
&&\left.+\sum_{\stackrel{\scriptstyle p=0}
{m\geq2}}^m\,2^{m-2}\,\frac{(2m-3)!!}{p!(m-p)!}
\frac{{\tilde S}_i^p}{(m+p-2)!}\frac{\pa^{m+p-2}}
{\pa x^{m+p-2}}\frac{\log(x+m_f)}
{R_i(x)^{m-1}}\left(\sum_{j\neq i}\frac{{\tilde S}_j}{x-e_j}\right)^{m-p}
\right|_{x=e_i}\,.\label{eq3}
\eea}
Again with the help of (\ref{s2}) it is possible to rewrite (\ref{eq3}) in the form 
(\ref{efe1}), and extract the form of the coefficients $\CF_q^{\chi=1}$. 
For example $\CF_2^{\chi=1}$ and $\CF_3^{\chi=1}$ are given by
{\small\bea
&&\CF_2^{\chi=1}=-\sum_{f=1}^{N_f}\sum_{i=1}^n\left(\frac{S_i^2}{2R_ie_{if}^2}
+\frac{S_i^2\,R_i'}{R_i^2e_{if}}\right)+\sum_{f=1}^{N_f}\sum_{i=1}^n\frac{2S_i}
{R_ie_{if}}\sum_{j\neq i}\frac{S_j}{e_{ij}}\,,\nonumber\\
&&\CF_3^{\chi=1}=\sum_{f=1}^{N_f}\sum_{i=1}^n\frac{S_i^3}{R_ie_{if}}\left(\frac{5R_i'R_i''}
{R_i^3}-\frac{6(R_i')^3}{R_i^4}-\frac{2R_i'''}{3R_i^2}-\frac{2(R_i')^2}
{R_i^3e_{if}}+\frac{R_i''}{2R_i^2e_{if}}-\frac{4R_i'}{3R_i^2e_{if}^2}+
\frac{1}{2R_ie_{if}^3}-\right.\nonumber\\
&&\left.
-2\sum_{j\neq i}\frac{1}{e_{ij}^3R_j}\right)+\frac{S_i^2}{R_ie_{if}}\left(\frac{8}{R_ie_{if}^2}
\sum_{j\neq i}\frac{S_j}{e_{ij}}+\frac{4R_i'}{R_i^2e_{if}}
\sum_{j\neq i}\frac{S_j}{e_{ij}}+\frac{12(R_i')^2}{R_i^3}
\sum_{j\neq i}\frac{S_j}{e_{ij}}-\frac{5R_i''}{R_i^2}\sum_{j\neq i}\frac{S_j}{e_{ij}}
\right.-\nonumber\\
&&\left.-\frac{4}{e_{if}^2}\sum_{j\neq i}\frac{S_jR_j'}{R_je_{ij}^2}+\frac{1}{R_ie_{if}}\sum_{j\neq i}
\frac{S_j}{e_{ij}^2}+\frac{8R_i'}{R_i^2}\sum_{j\neq i}\frac{S_j}{e_{ij}^2}+
\frac{4}{e_{if}^2}\sum_{j\neq i}\frac{S_j}{R_je_{ij}^3}\right)-
\frac{S_i}{R_ie_{if}}\left(\frac{4}{R_ie_{if}}\sum_{\stackrel{\scriptstyle j\neq i}{\scriptstyle k\neq j,i}}
\frac{S_jS_k}{e_{ij}e_{ik}}+\right.\nonumber\\
&&\left.+\frac{16R_i'}{R_i^2}\sum_{\stackrel{\scriptstyle j\neq i}{\scriptstyle k\neq j,i}}
\frac{S_jS_k}{e_{ij}e_{ik}}+
\frac{8}{R_i}\sum_{\stackrel{\scriptstyle j\neq i}{\scriptstyle k\neq j,i}}\frac{S_jS_k}{e_{ij}e_{ik}^2}+
\frac{12}{R_i}\sum_{\stackrel{\scriptstyle j\neq i}{\scriptstyle k\neq j,i}}\frac{S_jS_k}{e_{ij}^2e_{ik}}-
\sum_{\stackrel{\scriptstyle j\neq i}{\scriptstyle k\neq j,i}}\frac{4R_j'S_jS_k}{R_j^2e_{ij}e_{jk}}-
\sum_{\stackrel{\scriptstyle j\neq i}{\scriptstyle k\neq j,i}}
\frac{4S_jS_k}{R_je_{ij}e_{jk}^2}\right)
\eea}
where $e_{if}=e_i-m_f$. The result for $\CF_2^{\chi=1}$ agrees with the 
one computed in \cite{howard} perturbatively. As in the previous case, 
also note that if we set $S_{j\neq i}=0$ we recover the results computed in 
the previous section for the coefficients of order two and three in $S$. Notice that in 
this general case, as well as in the confining vacua case, the dependence of 
$\CF_{\chi=1}$ on the flavors is additive.

\section{Conclusions}

In this paper we have considered the Dijkgraaf--Vafa approach to $\CN=1$ 
supersymmetric theories with $SU(N_c)$ gauge group and $N_f<N_c$ matter 
hypermultiplets in the fundamental representation of the gauge group. 
Using this approach one can write the $\CF_{\chi=2}$ and $\CF_{\chi=1}$ 
contributions to the effective superpotential in terms of traces of 
certain matrix model operators (see Eq.(\ref{trw}) and Eq.(\ref{f1})). 
We found that those traces should be actually computed in the pure gauge theory even 
to calculate the effective superpotential for theories with flavor.
This remarkable fact allow us to compute the effective superpotential for theories with 
fundamental matter in a simple way without the need of performing 
a perturbative matrix model calculation, and just by computing quantities in the 
pure gauge theory.

Furthermore, we find that at the point of the moduli space where a 
maximal number of monopoles become massless Eq.(\ref{trw}) and Eq.(\ref{f1}) 
allow us to compute the exact effective superpotential for theories 
with fundamental matter in a simple way. This maximally 
degenerating point correspond to the case where the underlying Seiberg--Witten curve 
factorizes completely. As we already mentioned, the moduli that factorizes 
the Seiberg--Witten curve in the pure (no matter) case are well 
known and are given by the simple expression (\ref{mod}). Even though we are 
considering theories with fundamental matter, Eq.(\ref{trw}) and (\ref{f1}) 
tell us that those moduli are the only ingredient that we need 
to compute the exact effective superpotential in this case. 
We find that the result for the exact superpotential in this case is 
remarkably simple. 

We also considered the generalization of these results 
to an arbitrary point of the moduli space with $n$ distinct glueball superfields. 
We found that Eq.(\ref{trw}) and (\ref{f1}) also  
allow us to compute the effective superpotential for theories with 
fundamental matter in that case. To compute the superpotential we just 
needed the information about the spectral 
curve $y$ associated with the matrix model, $y=\sqrt{W'(x)^2+f_{n-1}(x)}$,
and the techniques developed in \cite{witten} to compute traces 
within the matrix model setup. The method developed in this paper provides 
a powerful tool to compute effective 
superpotentials avoiding perturbative matrix model calculations.

\section*{Acknowledgments}

The author would like to thank S. Naculich and H. Schnitzer for very 
useful conversations and for the critical reading of the manuscript. 
This work is supported by the DOE under grant DE--FG02--92ER40706.


\begin{thebibliography}{99}

\bibitem{civ}
F.~Cachazo, K.~Intriligator, and C.~Vafa, ``A large $N$ duality via a geometric transition,''
Nucl.\ Phys.\ {\bf B603} (2001) 3, {\tt hep-th/0103067}.

\bibitem{cv} F.~Cachazo and C.~Vafa,
``$\cN=1$ and $\cN=2$ geometry from fluxes,'' {\tt hep-th/0206017}.

\bibitem{dv} 
R.~Dijkgraaf and C.~Vafa,
``Matrix models, topological strings, and supersymmetric gauge theories,''
Nucl.\ Phys.\ {\bf B644} (2002) 3, {\tt hep-th/0206255}.

\bibitem{dv2} 
R.~Dijkgraaf and C.~Vafa,
``On geometry and matrix models,''
Nucl.\ Phys.\  {\bf B644} (2002) 21, {\tt hep-th/0207106}.

\bibitem{dv3} 
R.~Dijkgraaf and C.~Vafa, 
``A perturbative window into non-perturbative physics,'' 
Phys.\ Lett.\ B {\bf 573}, 138 (2003), {\tt hep-th/0208048}.

\bibitem{c}
R.~Dijkgraaf, M.~T.~Grisaru, C.~S.~Lam, C.~Vafa and D.~Zanon,
``Perturbative computation of glueball superpotentials,''
Phys.\ Lett.\ B {\bf 573}, 138 (2003), {\tt hep-th/0211017}.

\bibitem{witten}
F.~Cachazo, M.~R.~Douglas, N.~Seiberg and E.~Witten,
``Chiral rings and anomalies in supersymmetric gauge theory,''
JHEP {\bf 0212}, 071 (2002), {\tt hep-th/0211170}. 

\bibitem{seiberg}
N.~Seiberg,
``Adding fundamental matter to 'Chiral rings and anomalies in supersymmetric
gauge theory',''
JHEP {\bf 0301}, 061 (2003), {\tt hep-th/0212225}.

\bibitem{witten2}
F.~Cachazo, N.~Seiberg and E.~Witten,
``Phases of N = 1 supersymmetric gauge theories and matrices,''
JHEP {\bf 0302}, 042 (2003), {\tt hep-th/0301006}.
 
F.~Cachazo, N.~Seiberg and E.~Witten,
``Chiral Rings and Phases of Supersymmetric Gauge Theories,''
JHEP {\bf 0304}, 018 (2003), {\tt hep-th/0303207}.

\bibitem{arg} R.~Argurio, V.~L.~Campos, G.~Ferretti and R.~Heise,
``Exact superpotentials for theories with flavors via a matrix integral,''
Phys.\ Rev.\ D {\bf 67}, 065005 (2003), {\tt hep-th/0210291}.

\bibitem{m1} J.~McGreevy,
``Adding flavor to Dijkgraaf-Vafa,'' JHEP {\bf 0301}, 047 (2003), 
{\tt hep-th/0211009}.

\bibitem{m2}I.~Bena and R.~Roiban,
``Exact superpotentials in $\cN=1$ theories with flavor and 
their matrix model formulation,'' 
Phys.\ Lett.\ B {\bf 555}, 117 (2003), {\tt hep-th/0211075}.

\bibitem{m3} Y.~Demasure and R.~A.~Janik,
``Effective matter superpotentials from Wishart random matrices,''
Phys.\ Lett.\ B {\bf 553}, 105 (2003), {\tt hep-th/0211082}.

\bibitem{howard2}
S.~G.~Naculich, H.~J.~Schnitzer and N.~Wyllard,
``Matrix model approach to the N = 2 U(N) gauge theory with matter in the
fundamental representation,''
JHEP {\bf 0301}, 015 (2003), {\tt hep-th/0211254}.

\bibitem{hof}
C.~Hofman,
``Super Yang-Mills with flavors from large N(f) matrix models,''
JHEP {\bf 0310}, 022 (2003), {\tt hep-th/0212095}.

\bibitem{gukov}
R.~Dijkgraaf, S.~Gukov, V.~A.~Kazakov and C.~Vafa,
``Perturbative analysis of gauged matrix models,''
Phys.\ Rev.\ D {\bf 68}, 045007 (2003), {\tt hep-th/0210238}.

\bibitem{fer} F.~Ferrari,
``On exact superpotentials in confining vacua,'' Nucl.\ Phys.\ B {\bf 648}, 161 (2003), 
{\tt hep-th/0210135}.

\bibitem{fer2} F.~Ferrari,
``Quantum parameter space and double scaling limits in $\cN=1$ super 
Yang-Mills theory,'' Phys.\ Rev.\ D {\bf 67}, 085013 (2003) {\tt hep-th/0211069}; 
F.~Ferrari, ``Quantum parameter space in super Yang-Mills. II,''
Phys.\ Lett.\ B {\bf 557}, 290 (2003), {\tt hep-th/0301157}.

\bibitem{im2}
H.~Itoyama and A.~Morozov,
``The Dijkgraaf-Vafa prepotential in the context of general Seiberg-Witten
theory,''
Nucl.\ Phys.\ B {\bf 657}, 53 (2003), {\tt hep-th/0211245}.

\bibitem{howard}
S.~G.~Naculich, H.~J.~Schnitzer and N.~Wyllard,
``The N = 2 U(N) gauge theory prepotential and periods from a perturbative
matrix model calculation,''
Nucl.\ Phys.\ B {\bf 651}, 106 (2003), {\tt hep-th/0211123}.

\bibitem{yo}
M.~Gomez-Reino, S.~G.~Naculich and H.~J.~Schnitzer,
``Improved matrix-model calculation of the N = 2 prepotential,''
JHEP {\bf 0404}, 033 (2004), {\tt hep-th/0403129}.

\bibitem{sw}
N.~Seiberg and E.~Witten,
``Electric-magnetic duality, monopole condensation, and 
confinement in $\cN=2$ supersymmetric Yang-Mills theory,''
Nucl.\ Phys.\ {\bf B426} (1994) 19
[Erratum-ibid.\ {\bf B430} (1994) 485], {\tt hep-th/9407087}.

N.~Seiberg and E.~Witten, ``Monopoles, duality and chiral symmetry breaking 
in $\cN=2$ supersymmetric QCD,''
Nucl.\ Phys.\ {\bf B431} (1994) 484, {\tt hep-th/9408099}.

\bibitem{ds}
M.~R.~Douglas and S.~H.~Shenker,
``Dynamics of $\SU(N)$ supersymmetric gauge theory,''
Nucl.\ Phys.\ {\bf B447} (1995) 271, {\tt hep-th/9503163}.

\bibitem{pestun}
A.~Dymarsky and V.~Pestun,
``On the property of Cachazo-Intriligator-Vafa prepotential at the  extremum of
the superpotential,''
Phys.\ Rev.\ D {\bf 67}, 125001 (2003), {\tt hep-th/0301135}.

\bibitem{matone}
M.~Matone,
``Seiberg-Witten duality in Dijkgraaf-Vafa theory,''
Nucl.\ Phys.\ B {\bf 656}, 78 (2003), {\tt hep-th/0212253}. M.~Matone 
and L.~Mazzucato,
``Branched matrix models and the scales of supersymmetric gauge theories,''
JHEP {\bf 0307}, 015 (2003), {\tt hep-th/0305225}.

\bibitem{vy}
G.~Veneziano and S.~Yankielowicz,
``An Effective Lagrangian For The Pure N=1 Supersymmetric Yang-Mills Theory,''
Phys.\ Lett.\ B {\bf 113}, 231 (1982).

\bibitem{ils}
K.~A.~Intriligator, R.~G.~Leigh and N.~Seiberg,
``Exact superpotentials in four-dimensions,''
Phys.\ Rev.\ D {\bf 50}, 1092 (1994), {\tt hep-th/9403198}.

\bibitem{ook} H.~Fuji and Y.~Ookouchi,
``Comments on effective superpotentials via matrix models,''
 {\tt hep-th/0210148}.

\bibitem{georgi}
G.~Hailu and H.~Georgi,
``On exact superpotentials, free energies and matrix models,''
JHEP {\bf 0402}, 038 (2004), {\tt hep-th/0401101}.

\bibitem{janik}
Y.~Demasure and R.~A.~Janik,
``Explicit factorization of Seiberg-Witten curves with matter from random
matrix models,''
Nucl.\ Phys.\ B {\bf 661}, 153 (2003), 
{\tt hep-th/0212212}.

\bibitem{janik2}
R.~A.~Janik,
``Exact U(N(c)) $\to$ U(N(1)) x U(N(2)) factorization of Seiberg-Witten curves
and N = 1 vacua,''
Phys.\ Rev.\ D {\bf 69}, 085010 (2004), 
{\tt hep-th/0311093}.

\bibitem{warner}
K.~D.~Kennaway and N.~P.~Warner,
``Effective superpotentials, geometry and integrable systems,'' 
{\tt hep-th/0312077}.

\bibitem{gripaios}
B.~M.~Gripaios,
``Superpotentials for gauge and conformal supergravity backgrounds,''
 {\tt hep-th/0311025}. B.~M.~Gripaios and J.~F.~Wheater,
``Veneziano-Yankielowicz superpotential terms in N = 1 SUSY gauge theories,''
Phys.\ Lett.\ B {\bf 587}, 150 (2004), {\tt hep-th/0307176}.

\bibitem{planar}
E.~Brezin, C.~Itzykson, G.~Parisi and J.~B.~Zuber,
``Planar Diagrams,''
Commun.\ Math.\ Phys.\  {\bf 59}, 35 (1978).

\bibitem{SUN}
A.~Klemm, W.~Lerche, S.~Yankielowicz and S.~Theisen,
``Simple singularities and $\cN=2$ supersymmetric Yang-Mills theory,''
Phys.\ Lett.\ {\bf B344}, 169 (1995), 
{\tt hep-th/9411048}; 

P.~C.~Argyres and A.~E.~Faraggi,
``The vacuum structure and spectrum of $\cN=2$ supersymmetric $\SU(n)$ 
gauge theory,''
Phys.\ Rev.\ Lett.\  {\bf 74}, 3931 (1995), 
{\tt hep-th/9411057}.

\bibitem{im}
H.~Itoyama and A.~Morozov,
``Calculating gluino condensate prepotential,''
Prog.\ Theor.\ Phys.\  {\bf 109}, 433 (2003), 
{\tt hep-th/0212032}.

H.~Itoyama and A.~Morozov,
``Gluino-condensate (CIV-DV) prepotential from its Whitham-time  derivatives,''
Int.\ J.\ Mod.\ Phys.\ A {\bf 18}, 5889 (2003), 
{\tt hep-th/0301136}.

\end{thebibliography}
\end{document}